\documentclass[preprint2]{aastex}
\usepackage{apjfonts}

\newcommand{\gae}{\mathrel{\raise .4ex\hbox{\rlap{$>$}\lower 1.2ex\hbox{$\sim$}}}}

\shorttitle{X-ray Image of 3C 273 Jet}
\shortauthors{Marshall et al.}

\begin{document}

\title{Structure of the X-ray Emission from the Jet of 3C 273}

\author{H.L. Marshall\altaffilmark{1},
D.E. Harris\altaffilmark{2},
J.P. Grimes\altaffilmark{2},
J.J. Drake\altaffilmark{2},
A. Fruscione\altaffilmark{2},
M. Juda\altaffilmark{2},
R.P. Kraft\altaffilmark{2},
S. Mathur\altaffilmark{3},
S.S. Murray\altaffilmark{2},
P.M. Ogle\altaffilmark{1},
D.O. Pease\altaffilmark{2},
D.A. Schwartz\altaffilmark{2},
A.L. Siemiginowska\altaffilmark{2},
S.D. Vrtilek\altaffilmark{2},
B.J. Wargelin\altaffilmark{2}}
\altaffiltext{1}{Center for Space Research, Massachusetts Institute of
	Technology}
\altaffiltext{2}{Harvard-Smithsonian Center for Astrophysics}
\altaffiltext{3}{Ohio State University}
\email{hermanm@space.mit.edu}
% \slugcomment{DRAFT, \today, hlm}

\begin{abstract}
We present images from five observations of the
quasar 3C 273 with the Chandra X-ray Observatory.
The jet has at least
four distinct features which are not resolved
in previous observations.
The first knot in the jet (A1) is very bright in X-rays.
Its X-ray spectrum is well fitted with a
power law with $\alpha = 0.60 \pm 0.05$ (where S$_{\nu}
\propto \nu^{-\alpha}$).
Combining this measurement with lower frequency data
shows that a pure synchrotron model
can fit the spectrum of this knot from 1.647 GHz to 5 keV
(over nine decades in energy) with $\alpha = 0.76 \pm 0.02$, similar
to the X-ray spectral slope.  Thus, we place
a lower limit on the total power radiated
by this knot of $1.5 \times 10^{43}$ erg/s; substantially
more power may be emitted in the hard X-ray and $\gamma$-ray bands.

Knot A2 is also detected and is somewhat
blended with knot B1.  Synchrotron emission may also
explain the X-ray emission but a spectral bend
is required near the optical band.
For knots A1 and B1, the X-ray flux dominates the emitted energy.
For the remaining optical knots (C through H), localized X-ray enhancements 
that might correspond to the optical features are not clearly resolved.
The position angle of the
jet ridge line follows the optical shape with distinct, aperiodic
excursions of $\pm 1$\arcdeg\ from a median value of -138.0\arcdeg.
Finally, we find X-ray emission from the ``inner jet''
between 5 and 10\arcsec\ from the core.

\end{abstract}


\section{Introduction}

Previous high resolution observations of the 3C 273 jet using
Merlin at 1.647 GHz and
the Hubble Space Telescope (HST) Wide Field Planetary Camera 2
\citep{bahcall} showed that the overall shape of the
jet is somewhat different between the optical and radio bands.
The optical image appears dominated by elongated knots
roughly 0.1\arcsec\ by 0.5\arcsec\ in size, while the radio
image gives an indication of a ``cocoon'' structure, especially
at the end, or ``head''.  
Bahcall et al.\ speculated that the cocoon
is slowly moving material enveloping a relativistically moving flow.
Although relativistic motion is required to explain superluminal
motion in the quasar core, it is not yet clear that the
flow is relativistic in the jet on a kpc scale.

\begin{figure*}
\vspace{3in}
{\Large See color image: {\tt 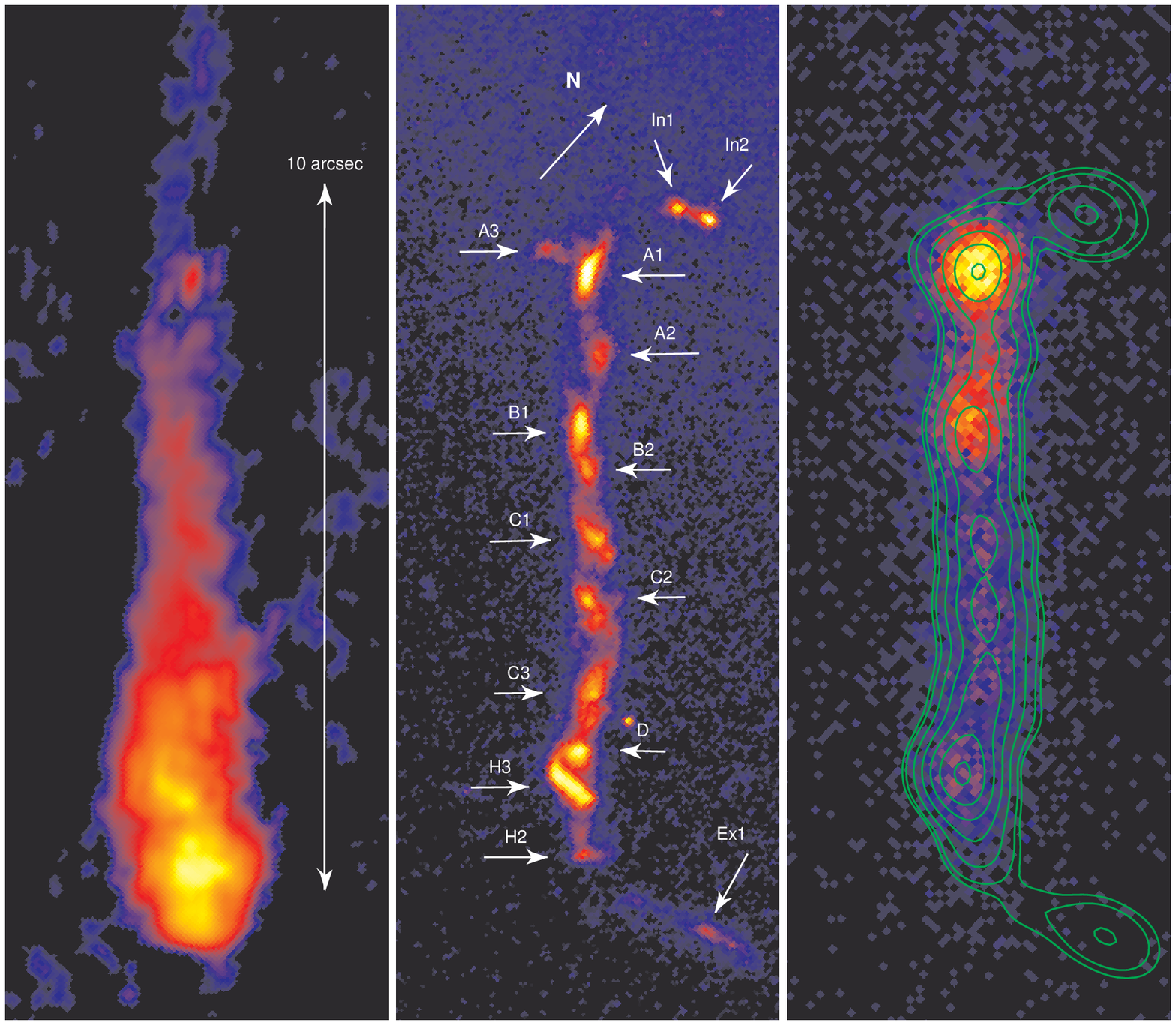}.}
\vspace{3in}
\caption{
Images of the jet in 3C~273 in three different bands.
{\em Left:} Image at 1.647 GHz using the Merlin array, kindly
provided by Tom Muxlow of Jodrell Bank.
{\em Middle:} The Hubble Space
Telescope Planetary Camera image in the
F622W filter (centered at 6170 \AA).
Features are labelled according to the nomenclature
used by \citet{bahcall}.
{\em Right:} Raw Chandra image of the
X-ray emission from the jet of 3C 273 in 0.1\arcsec\ bins
overlaid with a version of the HST image smoothed with a
Gaussian profile in order to match the X-ray imaging resolution.
The X-ray and optical images have been registered to each other
to about 0.05\arcsec\ using the position of knot A1.
The overall shape
of the jet is remarkably similar in length and curvature but
the X-ray emission fades to the end of the 
jet so individual C knots are not discernable.
Other differences are more apparent in Fig.~\ref{fig:profile}.
The radio emission is much fainter at knot A1 and is
displayed with a logarithmic scaling.}
\label{fig:image}
\end{figure*}

\citet{harris} used the {\em Einstein} Observatory data on 3C 273
to detect X-ray emission from the jet,
which was less than 1\% of the flux of the core.
More recently,
\citet{roeser}, examined ROSAT HRI images at $\sim$ 5\arcsec\
resolution to show that the X-ray emission drops with distance along
the jet.  Using models of the profile along the jet and
multi-color ground-based images at 1.3\arcsec\ resolution,
they generated spectral energy distributions (SEDs)
for knots in the jet and found that the X-ray flux of knot
A1\footnote{We will refer to the
knot naming convention given by \citet{bahcall} (reproduced
in Fig.~\ref{fig:image}).}
is consistent with that expected by extrapolating a simple synchrotron
model from the radio through the optical with one population of
electrons.  The highest energy electrons in their
model had $\gamma > 10^7$.
Synchrotron self-Compton (SSC) calculations generally fail to
predict X-ray intensities commensurate with those observed for any
of the knots.

X-ray images with the Chandra X-ray Observatory are now resolving
the spatial structure along quasar jets.  The first such image,
of the quasar PKS~0637--752, proved remarkable because of the
strong X-ray fluxes of the jet knots, relative to the radio fluxes
\citep{schwartz}.  Simple synchrotron and thermal models
could be ruled out easily while SSC
models required unreasonable conditions.
\citet{tavecchio} and \cite{celotti} have suggested
that inverse Compton scattering of the cosmic microwave background
could produce the required X-ray fluxes and require the
jet material to be moving relativistically at a small angle
to the line of sight as in models of the core.
We present Chandra images and spectra for
the jet in 3C 273 that we examine in light of these models.

\section{Observations and Analysis}

\subsection{X-ray Data Reduction}

3C 273 was observed three times as part of the calibration of the
Chandra grating spectrometers, once in a
direct imaging mode as part of science verification, and twice
more as the gratings failed to insert during calibration observations.
See Table~\ref{tab-3c} for a list of observations, totalling
193,230 s of exposure time.
Results from the dispersed spectra of the core are still being
analyzed as part of the ongoing effort to verify the grating
spectrometer effective area calibration and will be presented elsewhere.
The Low Energy Transmission Grating (LETG) and High Resolution Camera
(HRC) combination gave significantly fewer counts in the
jet compared to the others, so was not used in the combined X-ray
image.
We present here the
results from the zeroth order portion of the ACIS grating
observations and combine these data with that of the HRC-I and ACIS-S
imaging observations for image analysis.
The dispersed spectra of the jet were extremely faint; we
determined that there were no significant emission lines and did
not examine these data any further.

Each of the grating observations had well-known artifacts
that distort the zeroth order images.  Two artifacts are relevant
to observations of the jet.  Bright sources
observed with ACIS will show up with
a ``readout streak'' (see the Chandra Proposers' Observatory
Guide, or POG).  In
no ACIS observation, however, did the streak affect the image of
the jet.  The LETG has two support structures that produce
diffraction patterns (see the POG).  This
pattern is spaced at 60\arcdeg\ intervals and, again, did not
interfere with the image of the jet.
Individual observations have slightly uncertain absolute
pointing, so we repositioned each image separately.
The HRC and LETG/ACIS
observations were combined by referencing to the core.
Due to pileup,
the two non-grating ACIS observations and the 
High Energy Transmission Grating observations
had zeroth order images that were severely affected by pileup so
the first knot was used as a reference.
We estimate the uncertainty in this procedure gives a relative
offset between the core and the jet of
less than 0.05\arcsec\ by examining the location of
the center of the wings of the core
and comparing to the position of the first knot in each data set.

\begin{figure*}
%   \begin{minipage}[c]{0.5\textwidth}
  \includegraphics*{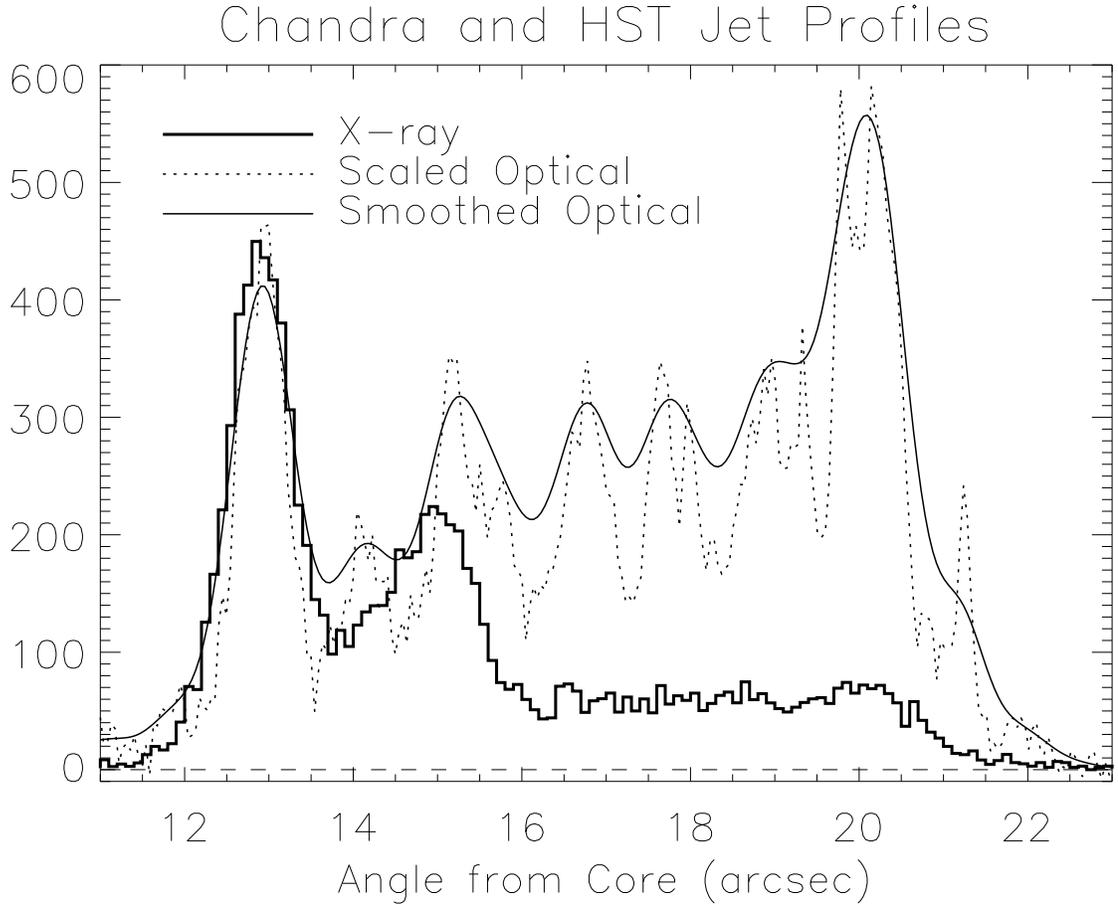}
\caption{
Profiles of the 3C 273 jet in the X-ray (histogram,
in counts per 0.1\arcsec\ bin) and optical bands.
The optical data are from a
Planetary Camera image in filter F622W, taken from
the Hubble Space Telescope (HST) archive,
are scaled to match the X-ray histogram for
knot A1: 0.325 $\mu$Jy per 0.04554\arcsec\ bin
at a vertical value of 500.
The raw optical profile was smoothed with a Gaussian
with $\sigma = 0.25$\arcsec\ (FWHM = 0.60\arcsec) and
was scaled to 0.27 $\mu$Jy per 0.04554\arcsec\ bin
at a vertical value of 500.
The optical profiles were displaced 0.22\arcsec\ closer to the
core to provide a better match between the X-ray and optical
profiles of knot A1; systematic registration uncertainties
are of this order.  Beyond knots A1 and B1 (12.9\arcsec\ and
15.0\arcsec\ from the core, respectively), other knots are
not clearly detected individually in the X-ray profile.}
\label{fig:profile}
% \end{minipage}
\end{figure*}

Figure~\ref{fig:image} shows the combined X-ray image binned
in 0.2\arcsec\ pixels.
The jet shows clear curvature but mostly has a one-dimensional
appearance, so a profile was computed for quantitative analysis.
The X-ray profile (Fig.~\ref{fig:profile})
was derived by summing data in a 1.5\arcsec\ wide
window centered on a position angle of -138.0\arcdeg, which is
the centerline of the jet.
We estimate that the correction for aperture losses is
10$\pm$ 5\% at any location along the jet, based on using a
wider extraction region.
The peak of emission occurs just under 13\arcsec\ from
the core.  Fitting a Gaussian profile to it, 
the centroid is 12{\farcs}93 $\pm$ 0{\farcs}01 from the core and the 
dispersion of a Gaussian is 0{\farcs}33 $\pm$ 0{\farcs}01 (for
a FWHM of 0.78\arcsec).  The $2\sigma$ limit
to the FWHM of this knot is 0{\farcs}3, given that point sources
have a projected FWHM of 0{\farcs}75, which is determined
from readout streak data for the core of 3C 273 and other
bright point sources.  The next peak in the
X-ray emission is clearly extended, from 14.0 to 15\arcsec\ from
the core, dropping more steeply on the downstream side.  From
16\arcsec\ to 21\arcsec\ the X-ray emission appears somewhat devoid
of distinct features but with some possible surface brightness
variations; there may be an unresolved knot at 20\arcsec.
The X-ray flux reaches the background level at 21\arcsec\ from
the nucleus.

Several regions were selected for X-ray spectral fitting: a) a
1\arcsec\ radius circle centered on the
first bright knot, b) a similar circle centered
on the extended knot at 15\arcsec\ and c) a rectangular box
extending from 16\arcsec\ to the end of the jet.
The 0.5-8 keV ACIS-S spectra were combined and fitted with power
law models holding $N_H$ fixed at $1.71 \times 10^{20}$ cm$^{-2}$
\citep{nh}.
The resultant spectral indices were
$\alpha_a = 0.60 \pm 0.05$,
$\alpha_b = 0.88 \pm 0.07$, and
$\alpha_c = 0.75 \pm 0.05$ (where $S_{\nu}
\propto \nu^{-\alpha}$); no spectral evolution is detected
along the jet.
Flux densities of several knot regions are given in Table~\ref{tab-fluxes}.
We estimate that the total jet power is about 0.4\% of the core power
in the 0.5-5.0 keV band.

\begin{figure*}
%   \begin{minipage}[c]{0.5\textwidth}
     \includegraphics*{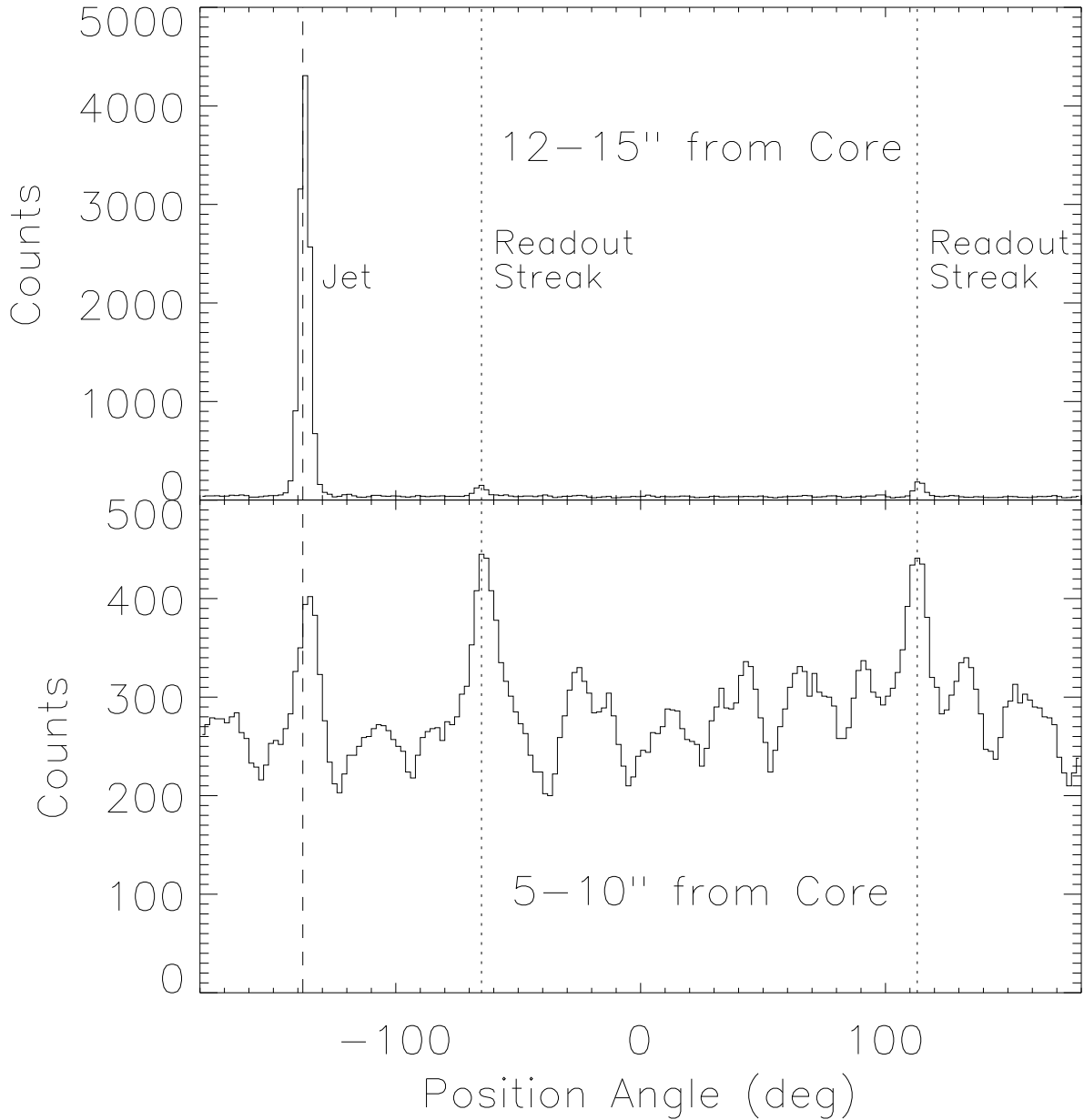}
\caption{
Azimuthal histograms from two different annuli centered
on the quasar core.  The top panel,
for the annulus from 12-15\arcsec\ from the core,
shows the jet at a position angle of -138\arcdeg\ (shown by
dashed lines) and the readout
streaks at about 110\arcdeg\ and -70\arcdeg\ (shown by dotted
lines).
The bottom panel shows the histogram for the 5-10\arcsec\ annulus.
The two brightest peaks are at the position angles of the readout
streak while the third brightest is at the same position angle as
the large scale jet.  The ripple is due the mirror support structure,
which causes shadows every 30\arcdeg.  The peak at -138\arcdeg\
is a detection of X-rays from the ``inner jet'', which
is not very bright optically but is detected in the radio band.
The position angle of the inner jet is rotated slightly to the
north compared to the 5-10\arcsec\ portion of the jet.  A similar
rotation is observed in the Merlin map (see figure~\ref{fig:image}).}
\label{fig:innerjet}
% \end{minipage}
\end{figure*}

X-ray emission is just detectable between the core and the knot A1,
as shown in Fig.~\ref{fig:innerjet}.
The 5-10\arcsec\ annulus shows a peak at the position angle of the
10-20\arcsec\ jet in addition to peaks at the position angles of
the readout streak.  It is difficult to quantify precisely the
flux of the inner jet due to the ripple inherent in azimuthal profiles
this close to a bright source which is caused by mirror support
structure.  Accounting for the ripple by fitting a sinusoid to 
the local background, we estimate that the count rate from the inner
jet is about 0.012 $\pm$ 0.001 count s$^{-1}$, corresponding
to a total flux density of 6.9 $\pm$ 0.6 $n$Jy at 1 keV for a power
law spectrum with $\alpha = 0.6$.

\subsection{Comparison to the Optical Emission}

X-ray components were identified using
images obtained from the HST archives.
Fluxes were determined in filters F450W, F622W, F814W, and
F160W (NICMOS).\footnote{Based on observations made with the
NASA/ESA Hubble Space Telescope, obtained from the data archive
at the Space Telescope Science Institute.  STScI is operated by
the Association of Universities for Research in Astronomy,
Inc. under NASA contract NAS 5-26555.}
The planetary camera (PC) observation using filter F622W
was used for direct comparisons to the X-ray jet profile, so the centroid
of the quasar was determined by isolating the image diffraction
spikes \citep{bahcall}, fitting these
with lines, and determining the intersection of the two lines (only two
spikes were available from this image).
Spatial distortions were corrected using polynomial coefficients
given by \citet{holtzman}.
The optical profile was obtained using the same method as used for
the X-ray profile and is also shown in Fig.~\ref{fig:profile}.
An offset of 0.22\arcsec\ was found between the centroids of the
the first X-ray peak and the peak of knot A1 from the PC image.
The positional uncertainty is
dominated by systematic uncertainties in measurement of the quasar
core, so this difference between the X-ray and optical positions
of this bright knot are not likely to be significant.  The optical emission
of A1 is clearly extended along a position angle closely aligned
to the overall PA of the jet (see Fig.~\ref{fig:image});
the profile is well fitted by a Gaussian with $\sigma =
0.3$\arcsec\ which is consistent with the X-ray profile at
the 2$\sigma$ level.

Identifying the sources of the remaining X-ray emission is not
quite so straightforward as for knot A1.  Region b (the extended
knot at 15\arcsec\ from the core) is not consistent
with a single point source at knot B1 but is likely to
be a blend of point sources at knots B1 and A2,
a somewhat weaker knot in the HST image.
There appears to be a discrete source of X-ray emission
near the positions of knots D and H3.
More data are needed to tell if knots besides these --
i.e., C1, C2, and C3 -- are also discrete sources of X-ray emission.

\section{Discussion}

\subsection{Morphology}

The overall shape of the
jet is quite similar in the optical and X-ray bands: there
is distinct curvature and the lengths are about the same.
The optical emission between the knots is $\gae 0.5$\arcsec\ wide
(Bahcall et al.\ 1995) and the
X-ray emission is marginally consistent with this level of broadening.
There are several important differences, however.
Knot A1 is much more prominent
in the X-ray data than in the optical image.
Furthermore, the X-ray jet fades along the jet while the
optical knots have similar brightnesses.  \citet{roeser}
also noted this difference.

The X-ray emission from
knot A1 is consistent with a point source but we cannot yet
exclude the possibility that it
is as extended as the optical emission along the jet axis.
The Merlin map shows that this knot is similarly
extended in the 1.647 GHz band.  The
radio emission of the other knots is quite difficult
to discern within the radio cocoon, giving rise to
the impression that bulk of the radio emission is physically
distinct from the optical and X-ray emission regions.
We find no significant X-ray emission from either extensions
(inner or outer, see Fig.~\ref{fig:image}), lending
support to the interpretation
that these are unrelated to the jet \citep{rm91}.

\subsection{Spectra}

\begin{figure*}
%   \begin{minipage}[c]{0.5\textwidth}
     \includegraphics*[width=6.5in]{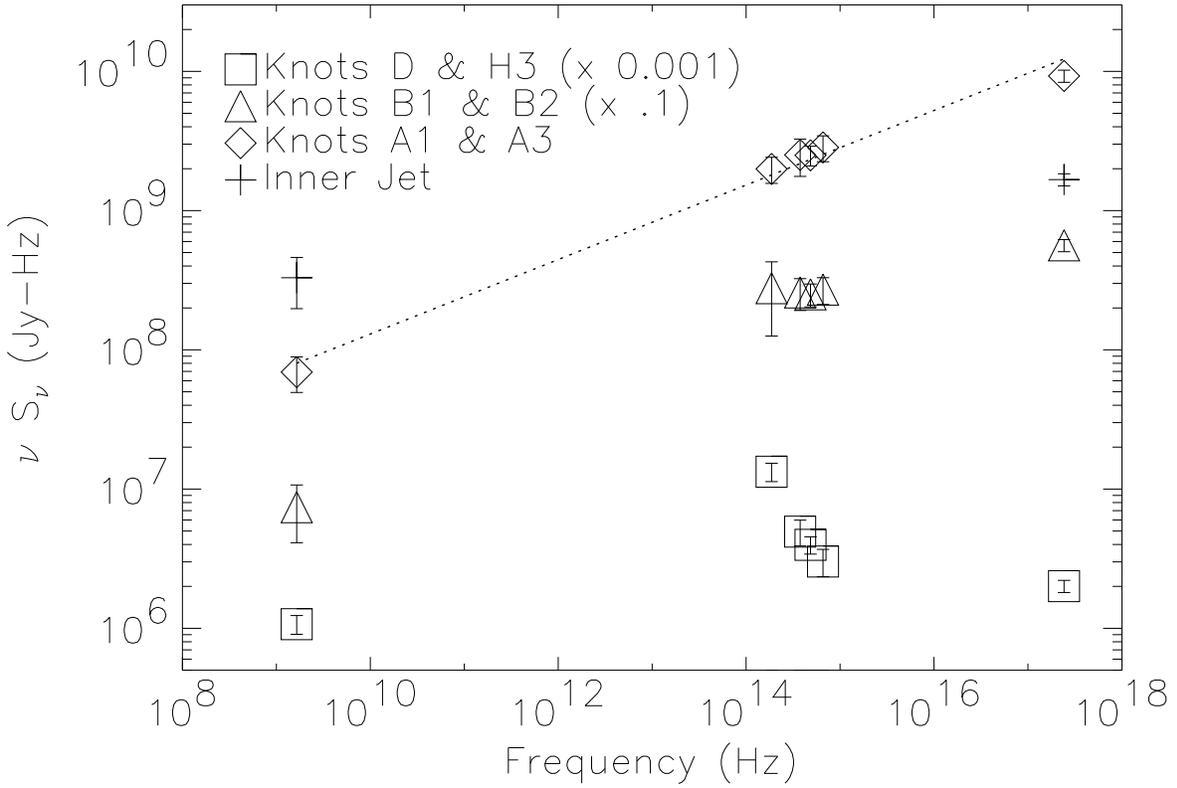}
\caption{
The spectral energy distributions (SEDs) of the 3C 273
jet knots A1, B1/B2, D/H3, and the inner jet.
The overall spectrum of knot A1 fits a simple
power law, shown as the dashed line with a slope of 0.25
so the spectrum is consistent with a simple synchrotron model with
a single power law distribution of electrons.
For knot B1, however, the SED appears to flatten between the radio
and optical bands; $\nu S_{\nu}$ still peaks in the X-ray band.
A single synchrotron model does not fit the SED of the D/H3
region.}
\label{fig:sed}
% \end{minipage}
\end{figure*}

The overall spectral energy distribution
(SED) for knot A1 (Fig.~\ref{fig:sed}) appears to fit a
simple synchrotron model, as suggested by \citet{roeser}.
Our estimate for the X-ray flux density is $\sim 2\times$ higher,
suggesting that the uncertainties derived from the ROSAT data
were underestimated.  Amazingly, the flux of this knot fits the
overall slope of the SED (based on table~\ref{tab-fluxes}),
0.76 $\pm$ 0.02, to
within the uncertainties, even after extrapolating over several
orders of magnitude in frequency.  Furthermore, the spectral slope
in the X-ray band is similar to that of the SED. 
Thus, the spectrum does not appear to break within the Chandra
bandpass, which is consistent with the excellent spectral fit
to the Chandra data.
The luminosity of knot A1, $1.5 \times 10^{43}$ erg/s
(for $q_0$ = 0.5 and $H_0$ = 70 km/s/Mpc), is about 40\% of the total X-ray
emission from jet, so
if the spectrum of knot A1 extends out to 100-200 keV with $\alpha$
= 0.6, then its X-ray flux will dominate the total power of the jet.
If the synchrotron break is above 5 keV then
$\gamma > 4 \times 10^7$
for the electrons, if one synchrotron model is to fit all the data.
The magnetic field is $80 \mu$G, based on minimum energy
arguments for nonrelativistic bulk motion.
For a cylindrical emitting volume of the size defined by the
optical emission, the SSC emission
from knot A1 would be less than 0.1 nJy, well below the observed
value.

Connecting the radio and optical/X-ray bands
of the B1 SED requires a slight bend,
consistent with the apparent flattening in the optical band.
While the form of the electron energy distribution may
not be a pure power law,
no spectral cutoff is observed in the SED, so the electron
energies may well reach energies comparable to those in knot A1.
R\"{o}ser et al.\ determined that SSC
models of knot B1 would not give rise to such large X-ray fluxes
and we confirm this conclusion.
As in knot A1, $\nu S_{\nu}$ is much higher in the X-ray band
than in the optical and radio bands.
Because $\alpha < 1$ in the X-ray band, we do not yet know where the
total powers of these knots peak.

Without detecting spectral cutoffs as observed in the first
knot of the jet in PKS~0637--752 \citep{schwartz}, we cannot
tell if there is a problem with the synchrotron model for knots
A1 and B1 as found in that source.
Similarly, the X-ray emission mechanism in the inner jet region is
difficult to model without spectra from the radio and optical bands.
Although there is no specific evidence that relativistic
motion is required to explain the X-ray fluxes of the
inner jet or knots A1 and B1,
as in the model suggested by \cite{tavecchio} and \cite{celotti},
this beaming model is a promising explanation for the X-ray
fluxes of the weaker knots so it could also provide a
an alternative to the synchrotron models for knots A1 and B1.
Thus, based on morphological similarities between
the X-ray and optical images and considering
that much of the jet power may be dissipated in the knots,
we speculate that the knots are
locations of internal shocks in a relativistic jet flow
that is decelerating before equilibrating with the ambient medium.
Alternatively, a helical jet structure would have regularly spaced
inflection points where the local jet flow is close
to the line of sight; beaming would be enhanced, giving
rise to to the observed knots.

\acknowledgments

We thank Herman-Josef R\"{o}ser for discussions about the
HST data for which he was the principal investigator.
We thank Tom Muxlow of Jodrell Bank for providing
the Merlin image.
This research is funded in part by NASA contracts NAS8-38249,
NAS8-39073, and SAO SV1-61010.

\begin{figure*}
%   \begin{minipage}[c]{0.5\textwidth}
     \includegraphics*[width=6.5in]{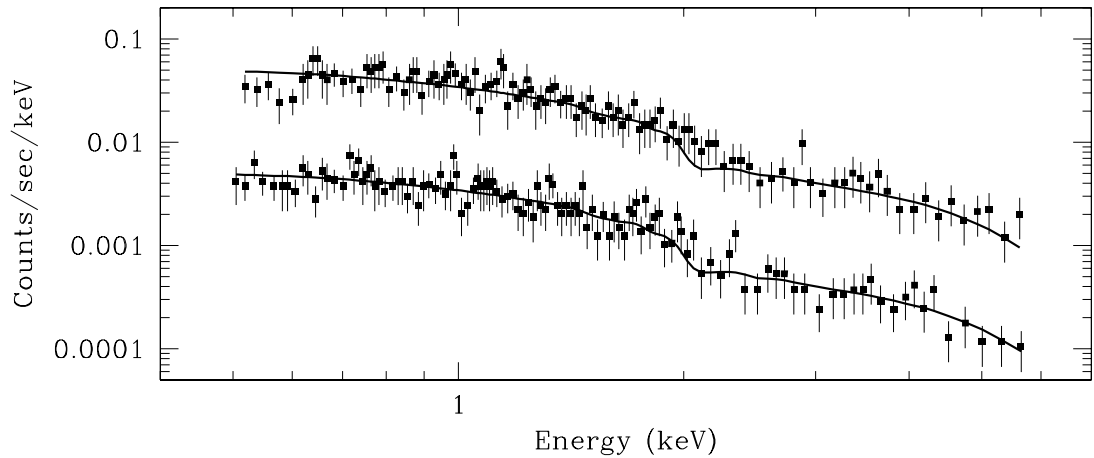}
\caption{
Two ACIS-S spectra that were simultaneously fit with a power law model
($\alpha = 0.60$; solid lines) including Galactic absorption of
$N_H=1.71\times10^{20}$ atoms/cm$^2$. Only data within 0.5-8~keV were included
in the analysis. The fitting was performed on the original 1024 PI bins using
Cash statistics in Sherpa. We rebinned the data to have minimum of 10 counts per
bin for the plotting purposes only. The upper data set, from observation
ID 1711, is displayed in the
original scale, while the lower data set, from observation
ID 1712, was rescaled downward by $\times 10$ for clarity.}
\label{fig:knotaspectrum}
% \end{minipage}
\end{figure*}

\clearpage

\begin{table}
\begin{center}
\caption{Chandra Observations of 3C 273}\label{tab-3c}
\begin{tabular}{lrcrc}
\tableline\tableline
Obs. ID & Detector & Grating & Date & Exposure \\
 & & & & (s) \\
\tableline
459 & ACIS-S & HETG & 10 Jan 2000 & 38600 \\
460 & HRC-I & none & 22 Jan 2000 & 20260 \\
461 & HRC-S & LETG & 9 Jan 2000 & 40280 \\
1198 & ACIS-S & LETG & 9--10 Jan 2000 & 38160 \\
1711 & ACIS-S & none & 14 Jun 2000 & 28130 \\
1712 & ACIS-I & none & 14 Jun 2000 & 27800 \\
\tableline
\end{tabular}
\end{center}
\end{table}

\begin{table}
\begin{center}
\caption{Fluxes of Knots in the 3C 273 Jet}\label{tab-fluxes}
\begin{tabular}{rccc}
\tableline\tableline
Frequency & \multicolumn{3}{c}{Flux Density} \\
(Hz) & \multicolumn{3}{c}{($\mu$Jy)} \\
 & A1 & B1 & D/H3 \\
\tableline
1.65 $\times$ 10$^9$ & 4.2 $\pm$ 1.2 $\times$ 10$^4$ &  
	4.5 $\pm$ 2.0 $\times$ 10$^5$ & 
	6.5 $\pm$ 1.0 $\times$ 10$^5$ \\
1.87 $\times$ 10$^{14}$ & 10.7 $\pm$ 2.3 & 14.8 $\pm$ 8.1 & 71.3 $\pm$ 10.7 \\
3.76 $\times$ 10$^{14}$ & 6.7 $\pm$ 2.0 & 6.9 $\pm$ 1.8 & 13.2 $\pm$ 2.8 \\
4.85 $\times$ 10$^{14}$ & 5.15 $\pm$ 0.83 & 5.16 $\pm$ 0.97 & 8.2 $\pm$ 1.2 \\
6.59 $\times$ 10$^{14}$ & 4.32 $\pm$ 0.91 & 4.11 $\pm$ 0.90 & 4.58 $\pm$ 1.02\\
2.42 $\times$ 10$^{17}$ & 0.038 $\pm$ 0.004 &  0.023 $\pm$ 0.002 & 
	0.0083 $\pm$ 0.0008 \\
\tableline
\end{tabular}
\end{center}
\end{table}

\end{document}